\documentclass[twocolumn,aps,prl,showpacs,floatfix,superscriptaddress]{revtex4}
\usepackage[T1]{fontenc}
\usepackage[latin9]{inputenc}
\usepackage{amsmath}
\usepackage{amssymb}
\usepackage{graphicx}
\usepackage{float}
\usepackage{siunitx}
\usepackage{xcolor}
\usepackage[english]{babel}

\usepackage{bm}

\makeatletter
\@ifundefined{textcolor}{}
{%
 \definecolor{BLACK}{gray}{0}
 \definecolor{WHITE}{gray}{1}
 \definecolor{RED}{rgb}{1,0,0}
 \definecolor{GREEN}{rgb}{0,1,0}
 \definecolor{BLUE}{rgb}{0,0,1}
 \definecolor{CYAN}{cmyk}{1,0,0,0}
 \definecolor{MAGENTA}{cmyk}{0,1,0,0}
 \definecolor{YELLOW}{cmyk}{0,0,1,0}
}

\@ifundefined{definecolor}{\usepackage{color}}{}

\usepackage{babel}

\usepackage{array}
\newcolumntype{L}[1]{>{\raggedright\let\newline\\\arraybackslash\hspace{0pt}}m{#1}}
\newcolumntype{C}[1]{>{\centering\let\newline\\\arraybackslash\hspace{0pt}}m{#1}}
\newcolumntype{R}[1]{>{\raggedleft\let\newline\\\arraybackslash\hspace{0pt}}m{#1}}

\makeatother

\begin{document}
\flushbottom

\title{Dynamics of a vortex lattice in an expanding polariton quantum fluid}
\author{Riccardo Panico}
\affiliation{Dipartimento di Matematica e Fisica E.~De Giorgi, Universit\`a del Salento, Campus Ecotekne, via Monteroni, Lecce 73100, Italy}
\affiliation{CNR NANOTEC, Institute of Nanotechnology, Via Monteroni, 73100 Lecce, Italy}
\author{Guido Macorini}
\affiliation{CNR NANOTEC, Institute of Nanotechnology, Via Monteroni, 73100 Lecce, Italy}
\author{Lorenzo Dominici}
\affiliation{CNR NANOTEC, Institute of Nanotechnology, Via Monteroni, 73100 Lecce, Italy}
\author{Antonio Gianfrate}
\affiliation{CNR NANOTEC, Institute of Nanotechnology, Via Monteroni, 73100 Lecce, Italy}
\author{Antonio Fieramosca}
\thanks{Present affiliation: Division of Physics and Applied Physics, School of Physical and Mathematical Sciences, Nanyang Techn. Univ., Singapore}
\affiliation{CNR NANOTEC, Institute of Nanotechnology, Via Monteroni, 73100 Lecce, Italy}
\author{Milena De Giorgi}
\affiliation{CNR NANOTEC, Institute of Nanotechnology, Via Monteroni, 73100 Lecce, Italy}
\author{Giuseppe Gigli}
\affiliation{Dipartimento di Matematica e Fisica E.~De Giorgi, Universit\`a del Salento, Campus Ecotekne, via Monteroni, Lecce 73100, Italy}
\affiliation{CNR NANOTEC, Institute of Nanotechnology, Via Monteroni, 73100 Lecce, Italy}
\author{Daniele Sanvitto}
\email{daniele.sanvitto@cnr.it}
\affiliation{CNR NANOTEC, Institute of Nanotechnology, Via Monteroni, 73100 Lecce, Italy}
\affiliation{INFN, Sez.Lecce, Via Monteroni, 73100 Lecce, Italy}
\author{Alessandra S. Lanotte}
\email{alessandrasabina.lanotte@cnr.it}
\affiliation{CNR NANOTEC, Institute of Nanotechnology, Via Monteroni, 73100 Lecce, Italy}
\affiliation{INFN, Sez.Lecce, Via Monteroni, 73100 Lecce, Italy}
\author{Dario Ballarini}
\affiliation{CNR NANOTEC, Institute of Nanotechnology, Via Monteroni, 73100 Lecce, Italy}

\begin{abstract}
If a quantum fluid is driven with enough angular momentum, at
equilibrium the ground state of the system is given by a lattice of
quantised vortices whose density is prescribed by the quantization of
circulation. We report on the first experimental study of the
Feynman-Onsager relation in a non-equilibrium polariton fluid, free to
expand and rotate. Upon initially imprinting a lattice of vortices in
the quantum fluid, we track the vortex core positions on picosecond
time scales. We observe an accelerated stretching of the lattice and
an outward bending of the linear trajectories of the vortices, due to
the repulsive polariton interactions. Access to the full density and
phase fields allows us to detect a small deviation from the
Feynman-Onsager rule in terms of a transverse velocity component, due
to the density gradient of the fluid envelope acting on the vortex
lattice.
\end{abstract}

\date{\today}

\maketitle

One of the most remarkable characteristics of a Bose-Einstein
condensate (BEC) is its response to rotation
\cite{RevFetter}. Differently from a conventional fluid, in the
``rotating bucket'' experiment a condensate does not rotate with the
bucket for angular velocities slower than a critical
value~\cite{Legget1999}. The absence of friction with the bucket walls
is a unique property of superfluids, which realize the ideal case of
irrotational flow. Yet, the velocity field of superfluids is
irrotational up to phase defects, i.e. quantized vortices, which allow
the condensate to have a finite angular momentum. As a consequence,
for a driving angular frequency larger than a critical
value~\cite{Dalibard2000}, the superfluid breaks into the formation of
quantized vortex filaments in 3D, or point-like vortices in 2D, as
observed in superfluid helium and ultracold atomic
condensates~\cite{RevModPhys1966,Barenghi_book}. More generally,
quantised vortices are excited states (topological defects) of a
quantum fluid which form also without macroscopic rotation of the
potential trap, for example via the Kibble Zurek mechanism or in
turbulent regimes~\cite{PRL2016,Barenghi_book2}. Quantised vortices
have also proven to be striking examples of the similarities between
the condensed matter, optical, and dilute-gas quantum systems, since
complex Ginzburg-Landau equations (CGLEs) describe a vast variety of
phenomena such as superconductivity, superfluidity, lasing and
Bose-Einstein condensation~\cite{Graham1970}. With respect to the
optical vortices observed in paraxial vortex beams, CGLEs include
light-matter interaction as a Kerr type
nonlinearity~\cite{Coullet1989, Allen1992}, allowing for the existence
of dark vortex solitons in a defocusing nonlinear medium and quantized
vortices in a superfluid~\cite{Desyatnikov2005,
  Carretero_book}.\\ \indent Exciton-polaritons (polaritons hereafter)
are a relatively new example of
superfluid~\cite{Carusotto2004,Amo2009,Lerario2017}, in which a
macroscopic coherent state is formed even far from the thermal
equilibrium condition~\cite{Caputo2018}. Polaritons are bosonic
quasi-particles which result from the strong interaction between light
and matter in semiconductor microcavities with embedded quantum
wells. In most cases, their dynamics is well described by a
generalized Gross-Pitaevskii (GP) equation, which takes into account
the driven-dissipative character of
polaritons~\cite{Carusotto2013}. \\ \indent In the past decade,
quantized vorticity in polariton fluids was observed under a variety
of pumping conditions~\cite{Lagoudakis2008,Tosi2012}. Highly nonlinear
effects on the nucleation of few vortices, and solitons have been
shown, as well as their all-optical manipulation and trapping in
propagating polariton fluids~\cite{Sanvitto2011, PhysLiew}. A major
advantage of this system is given by the photonic component, which
enables the control over the phase and density profiles of the
polariton fluid by optical shaping of the pumping laser beam
~\cite{Dominici2018, PhysHivet}. Additionally, the nonlinear
interactions inherited from the excitonic component are orders of
magnitude higher than in standard nonlinear media. High quality
samples now available, with longer polariton lifetime and reduced
density of defects allow to explore complex configurations of
vortices, going beyond previous realizations of a single or few
vortices.\\ \indent In this Letter, we report on the creation on
demand of a macroscopic lattice of quantised
vortices in polariton fluids and the measurement of the evolution of
both density and phase. The quantum fluid is free to expand and each
vortex has a dual function: it participates to the build up of the
rotation and it acts as a test particle that enables the observation
of the dynamics. We measure the lattice rotation and expansion, and
show that these exhibit a small but measurable deviation from the
Feynman-Onsager relation. In particular, we observe a detailed balance
between the faster radial separation of the vortex cores due to the
repulsive polariton-polariton interactions, and a slower rotation of
the quantum fluid, yet preserving the regular lattice shape. We model
these observations in terms of the initial vortex lattice density, or
equivalently its inter-vortex spacing, acting as the characteristic
scaling length which determines both the expansion rate of the lattice
and its instantaneous angular velocity. Finally, we highlight the role
of the gradients of quantum fluid density resulting in an additional
velocity contribution onto the rotation of the lattice, likewise a
Magnus effect of classical fluids.\\
\begin{figure}
    \centering
    \includegraphics[width=0.9\columnwidth]{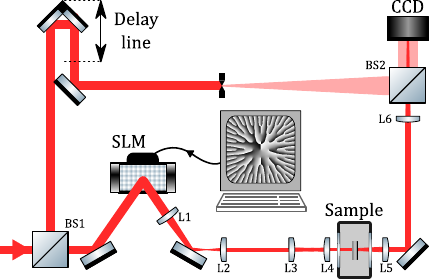}
    \caption{Schematic representation of the experimental setup. A
      laser beam is divided in two paths by a beam splitter (BS1) and
      one beam is diffracted by a SLM, where the phase pattern of a
      lattice of vortices is displayed. The beam with the imprinted
      phase profile is imaged on the sample surface by a system of
      lenses. The signal is made to interfere with a
        time-delayed reference beam. The interferogram is acquired by
      a CCD camera, and both density and phase are reconstructed in
      space and time, by digital off-axis holography.}
    \label{fig:setup}
\end{figure}
\begin{figure}
    \centering
    \includegraphics[width=0.9\columnwidth]{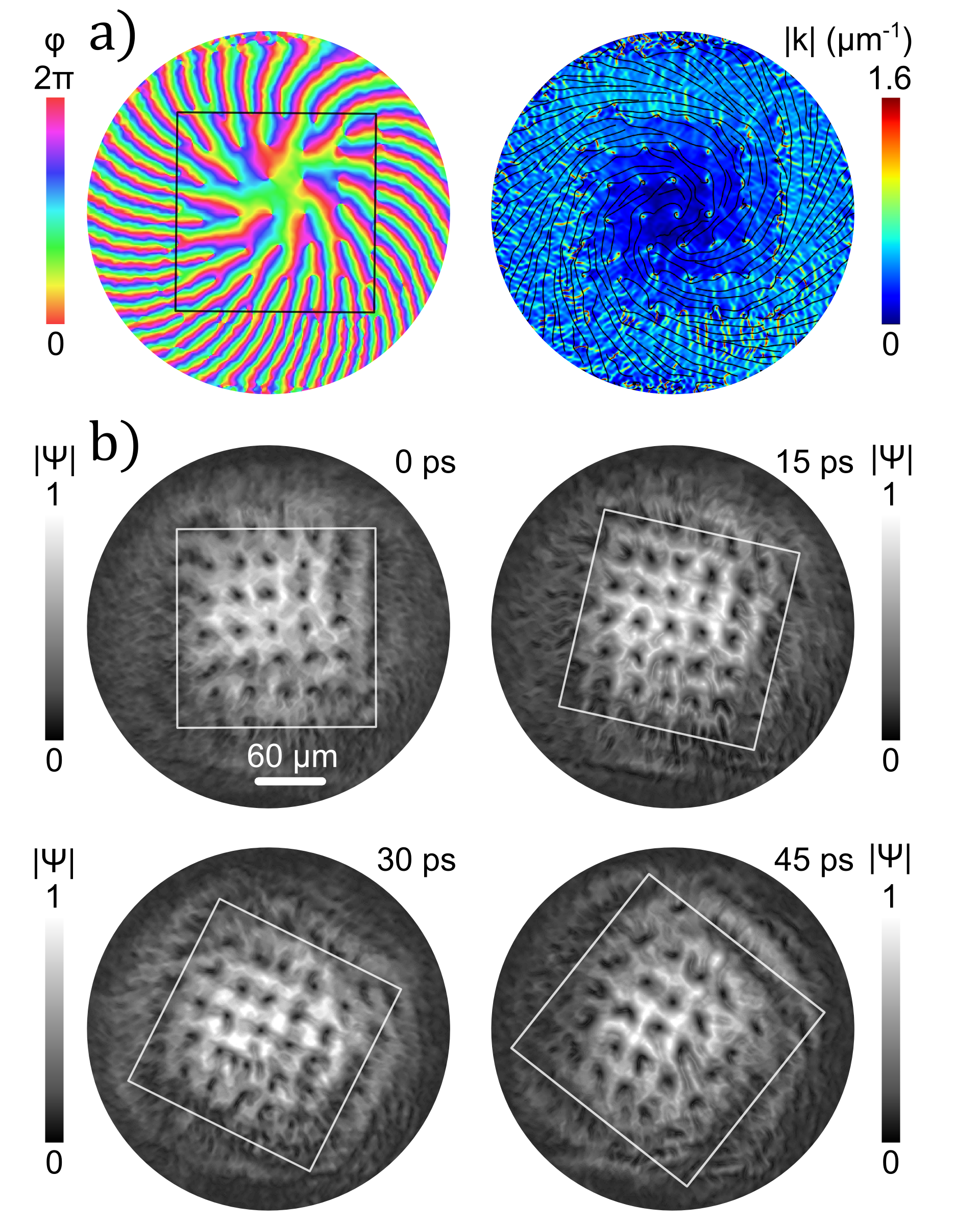} 
    \caption{(\textbf{a}) Phase map $\varphi(\textbf{r})$ of the
      initial state of a $7 \times 7$ square lattice of vortices
      (left) with momentum $\textbf{k}(\textbf{r}) ={\bm
        \nabla}\varphi$ representing the velocity field
      (right). The color scale of the momentum map is
        bounded at $|k| = 1.6$ $\mu$m$^{-1}$ to avoid saturation
        inside the cores. The regular lattice builds up a continuous
        increase of azimuthal momentum reaching a maximum at its
        boundary of $\approx \SI{0.42}{\micro\meter}^{-1}$,
        corresponding to a fluid velocity of
        $\SI{1.2}{\micro\meter}~\SI{}{\pico\second}^{-1}$. Overlapped
        streamlines (black lines) display the velocity direction,
        indicating the rotating motion. (\textbf{b}) The normalized
      amplitude maps over a span of 45 ps. The outer boundary of the
      lattice is shown by a square (thin line).}
    \label{fig:time}
\end{figure}
We use a semiconductor planar microcavity with 12 GaAs quantum wells
embedded in two distributed Bragg reflectors. A pulsed excitation
tuned on resonance with the polariton energy is used to imprint the
vortex lattice state~\cite{Dominici2018}. The phase wavefront of the
exciting beam is modulated by a Spatial Light Modulator (SLM),
consisting of an array of individually programmable pixels made of
liquid crystal cells. The phase profile of the vortex lattice is
designed by software, sent to the SLM and transferred to the pumping
beam upon reflection on the SLM screen. The SLM and the microcavity
are conjugate planes in the optical excitation path, with the image of
the vortex lattice reduced by a factor 50 in size on the sample
surface (see Fig.1). The exciton-polariton phase profile is inherited
from the pulsed laser upon resonant excitation and is free to evolve
after the pulse has gone (pulse duration of 2~ps). The time resolved
evolution of the vortex lattice is obtained by interfering the signal
coming from the microcavity with a sample of the exciting pulse, as
shown in Fig.~\ref{fig:setup}. Digital off-axis holography allows to
retrieve the spatial distribution of both density and phase of the
polariton fluid in the 2D plane of the
microcavity~\cite{Donati2016}. By changing the time delay between the
pump and the reference, the evolution dynamics of the 2D quantum fluid
is obtained in both space and time domain (see \cite{suppl} method
section).

The phase and the velocity fields of the initial state of the system
with a regular lattice of $7\times 7$ vortices with the same unit
topological charge are shown in Fig.~\ref{fig:time}a\,. Due to the
internal concentration of vortex charges, the largest momentum is
reached at the outer boundary of the region (apart from local
fluctuations). The background amplitude profile (Fig. 2b) of the
lattice is not uniform, but modulated by the Gaussian profile of the
laser beam. Once the pulsed resonant excitation is over, the polariton
lattice rotates in a rigid-body movement, such that the fluid,
irrotational for simply-connected regions, effectively appears as
rotational in a coarse grained picture (in Fig.~\ref{fig:time},
time-shots illustrate the evolution of the polariton density during
the first 45~ps).

The velocity circulation around a multiply connected region enclosing
a lattice of unitary vortices is quantised according to the
Feynman-Onsager relation~\cite{Feynman1955, Donnelly1993},
resulting in an angular velocity of the lattice
\begin{equation} 
\Omega = \frac{h}{2m}\frac{1}{d^2},
\label{eq:feynman}
\end{equation}
with $m$ the polariton mass and $d$ the intervortex distance~\cite{suppl}.

In experiments with superfluid helium \cite{Yarmchuk1979,Be2006}, when
the system is {\it put} into rotation at constant frequency $\Omega$,
at {\it equilibrium} a regular lattice of vortices of equal sign
unitary charge is formed with an average density in agreement with
Eq.~(\ref{eq:feynman}). Experiments with gaseous BECs in cylindrical
traps~\cite{Abo-Shaeer2001, Raman2001} confirm these results with the
formation of triangular (or hexagonal) lattices, which are ground
state configurations in the rotating frame, containing up to hundreds
of vortices. However, in our system, the polariton fluid expands due
to the absence of the confining potential and moreover a stationary
state can never be reached, leading to both the vortex spacing and the
rotation frequency to change in time. In order to quantitatively
describe the change of the inter-vortex distance, we can think to the
initial condition as that of a rigid-body rotation, in agreement to
the Feynman-Onsager relation, with the azimuthal velocity proportional
to the distance from the centre, ${v} = ({\bm \Omega} \times {\bm r})
\cdot \hat{\bm e}_{\theta}$. In the absence of interactions, given
that the fluid is free to expand, every particle continues to move
along a straight line with the initial velocity $\bm
v$~\cite{Rozas97}. The inter-vortex distance $\mathrm{d}(t)$ increases
following a law of analogue form to what is expected
for the density of a diffracting optical beam and for an expanding BEC
of non-interacting particles after the release from a magnetic trap,
\begin{equation}
d(t) = d_0 \sqrt{1+\left(\frac{a}{d_{0}^2}t\right)^2}.
\label{eq:dist}
\end{equation}
Here $d_{0}$ is the initial distance and $a = a_0 \equiv h/2m$ in the
linear regime.
\begin{figure}
    \includegraphics[width=0.85\columnwidth]{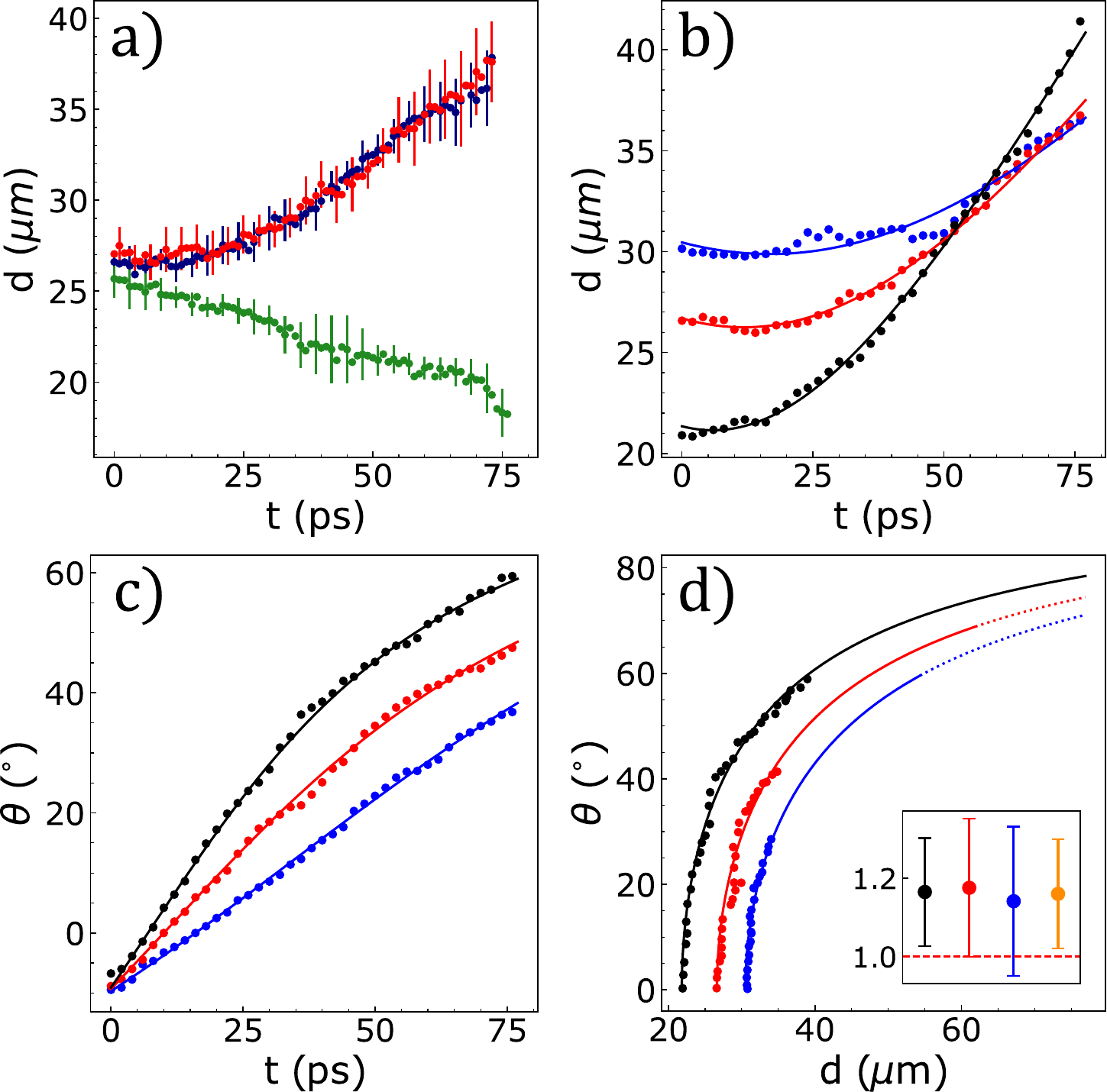}
    \caption{(\textbf{a}) Inter-vortex distances over time for three
      cases with total charge equal to $Q=\pm 49$ (red and blue dots,
      respectively), and $Q\simeq 0$ (green dots). Error bars come
      from the estimate of the vortex positions. (\textbf{b}) Mean
      inter-vortex distances for three different initial separations
      and $Q=-49$. (\textbf{c}) The orientation angle of the whole
      lattice, for the same cases as in (b). In panels (b) and (c),
      solid lines are the best fits of Eqs.~\ref{eq:dist} and
      \ref{eq:angle}, with $a=11.6\pm
      0.9\,\SI{}{\micro\meter\squared\pico\second^{-1}}$ and
      $b=0.99\pm 0.09$. The small contraction of the lattice at short
      time lags, due to a residual curvature of the phase profile of
      the exciting beam, is taken into account by introducing a time
      offset $t_0$ in the fitting functions used in panel (\textbf{b,
        c}): $t_0$=7, 11 and 17~$\SI{}{\pico\second}$ for $d_0=$21,
      26.5 and 30~$\SI{}{\micro\meter}$, respectively. (\textbf{d})
      Rotation angle as a function of vortex distance during the
      expansion of the lattice for the three cases shown in panels (b)
      and (c). Only positive angles are shown in panel (d), i.e. for
      $t>t_0$, to allow the comparison between different $d_0$. Solid
      lines correspond to the first $\SI{150}{\pico\second}$ of
      expansion and rotation. Inset: the value of $\frac{a\,b}{a_0}$,
      as extracted from individual fit (same color legend as in panels
      (b) and (c), and for a global fit (yellow point). The deviation
      from the Feynman-Onsager relation is quantified by the distance
      from the dashed line, with
      $a_0=9.85\,\SI{}{\micro\meter\squared\pico\second^{-1}}$ and
      $b=1$, of the linear case.}
    \label{fig:dist}
\end{figure}
The expansion factor $(a/d_0^2)$ sets the scaling of all the distances
in the lattice. Noteworthy, the circular symmetry of initial
velocities is such that any shape, such as the square lattice, appears
to be expanding and rotating at later times. This angle of rotation,
directly linked to the Gouy phase, can be written upon
geometrical considerations as:
\begin{equation}
\theta (t) = b \cdot \arctan{\left(\frac{a}{d_0^2}t\right)}\,,
\label{eq:angle}
\end{equation}
with the prefactor $b = 1$ in the linear regime, meaning that the
limit angle tends to $\pi/2$ for large times. If we add the repulsive
interactions between polaritons, the initial mean-field energy is
expected to be partially released into kinetic energy during the
expansion~\cite{Mewes1996, Holland1997, Schunck2007}.  Nonlinear
repulsion marks the origin of a dynamical regime where the vortex
trajectories deviate from the straight lines of the linear case due to
the earlier onset of an additional radial component of the
velocity. As a consequence, in the nonlinear case,
  the expansion factor in Eqs.~(\ref{eq:dist}) and ~(\ref{eq:angle})
  is expected to be always larger than in the linear limit,
  $a>a_{0}$.

The evolution of the inter-vortex distances in regular lattices of
same-sign vortices, and in a lattice of vortices and antivortices is
shown in Fig.~\ref{fig:dist}a. While the averaged spacing $\mathrm{d}$
increases independently of the sign of the circulation ($Q\neq 0$),
when the total injected topological charge is null (lattice of
vortices and antivortices), the rotation rate is zero~\cite{suppl} and
the inter-vortex spacing slightly shrinks due to the mutual attraction
of vortices with opposite sign. The time behaviour of the average
spacing $d$ and rotation angle $\mathrm{\theta}$, measured during the
expansion of lattices with different initial inter-vortex distance
$d_0$, are shown in Fig.~\ref{fig:dist}b and Fig.~\ref{fig:dist}c,
respectively. The solid lines are the best global fit of
Eqs.~(\ref{eq:dist}) and Eq.~(\ref{eq:angle}) to experimental data,
showing a very good agreement with a single set of parameters. In
Fig.~\ref{fig:dist}d, the rotation angle is shown as a function of the
intervortex separation for the same data reported in
Fig.~\ref{fig:dist}b-c. These results have been confirmed by
independent analysis of numerical simulations (see
\cite{suppl}). \\The rigid-like rotation of the lattice allows us to
compare the Feynman-Onsager relation in Eq.~(\ref{eq:feynman}) with
the measurements of the vortex trajectories. Indeed, from
Eqs.~(\ref{eq:dist}) and ~(\ref{eq:angle}), we obtain
\begin{equation}
    d^2(t)\frac{d\theta(t)}{dt}\,=\,a\,b.\notag
\end{equation}
Therefore, the angular velocity $\frac{d\theta(t)}{dt}$ is inversely
proportional to the squared intervortex distance $d(t)$ during the
whole expansion of the lattice and their product is the same for the
three initial $d_{0}$ shown in Fig.~\ref{fig:dist}. \\In the inset of
Fig.~\ref{fig:dist}d, the product $ab$ is compared to the equilibrium
value $a_0 = h/2m$ of Eq.~(\ref{eq:feynman}), showing a measurable
deviation from the Feynman-Onsager relation. The difference is small,
but can be appreciated for each separate $d_{0}$, as well as for the
global best fit over the three evolutions (yellow point). We ascribe
such deviation to the Magnus effect, i.e., the transversal velocity of
the vortex cores induced by density gradients in the polariton
fluid~\cite{Kivshar1998,Navarro2013,Simula2018}. In our experiments,
the density gradient (similar for the three initial $d_{0}$, since it
depends on the Gaussian envelope of the same pumping beam) points
radially inwards and the Magnus-like velocity accelerate the rotation
of the lattice with respect to that of the fluid, $a \,b > h/2m$.\\
\begin{figure}
    \centering
    \includegraphics[width=1.\columnwidth]{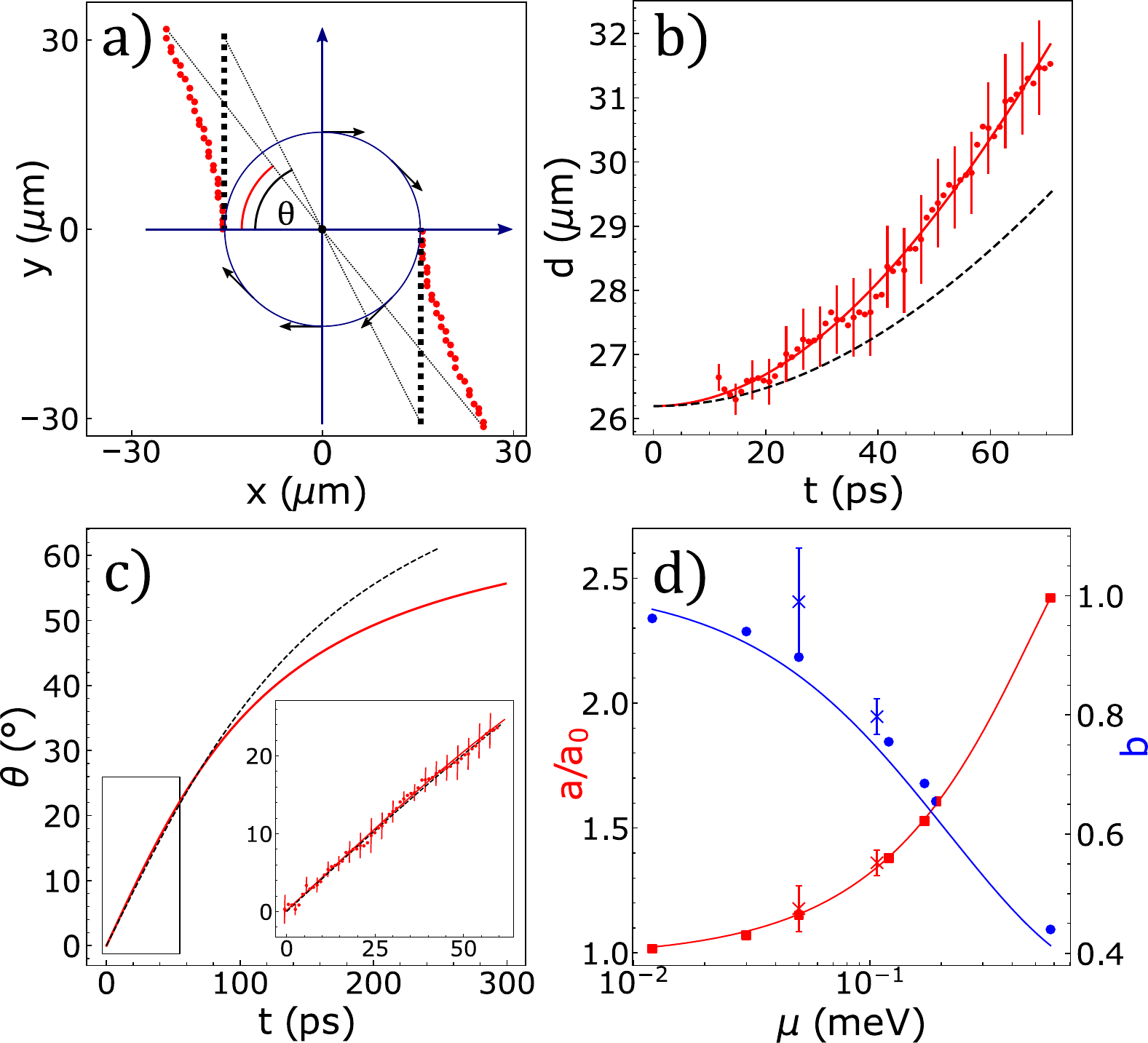} 
    \caption{\textbf{(a)} Graphical representation of the effect of
      nonlinear interactions, producing a bending of the trajectories
      from the straight lines. \textbf{(b)} The time evolution of the
      inter-vortex separation $\mathrm{d}$ for a lattice of $ 5\times
      5$ vortices. Red points with error bars are experimental data,
      at chemical potential $\mu=\SI{0.12}{\milli\electronvolt}$
      (corresponding to a polariton density $n \sim
      \SI{100}{\micro\meter^{-2}}$, with an interaction strength
      $g=10^{-3}\SI{}{\milli\electronvolt\micro\meter^2}$), and red
      solid line is the result of a numerical simulation at equal
      $\mu$; from the best fit, the corresponding nonlinear expansion
      factor is $a=1.38\, a_0$. The black dashed line is the evolution
      in the linear case, corresponding to $\mu \simeq
      \SI{0}{\milli\electronvolt}$, and
      $a=a_0=5.04\,\SI{}{\micro\meter\squared\pico\second^{-1}}$. \textbf{(c)}
      Rotation angles as a function of time corresponding to the
      lattice expansions shown in (b): red line is the best fit of
      Eq.~\ref{eq:angle} with $a=1.38\, a_0$ and
        $b=0.8$; the black dashed line is the angle evolution in the
      linear case ($a=a_0$, $b=1$); in the inset, a
      zoom at small time lag show the experimental data (red points)
      superposed to the best fit (red line). \textbf{(d)} Parameters
      ($a,b$) extracted from experiments (crosses) and simulations
      (circles, squares) at different values of $\mu$. The red line is
      a polynomial fit to $a$ values, the blue line is the expected
      behaviour for $b$ in the absence of the Magnus
      effect, $a_{0}/a$.}
    \label{fig:newexp}
\end{figure}
To highlight the role of nonlinearities in the dynamics, we move to a
position on the sample with a higher excitonic fraction (see
\cite{suppl}). In Fig.~\ref{fig:newexp}a, the trajectories of a vortex
pair in the polariton fluid are compared to the straight ones of the
linear case. The faster increase of the intervortex distance, with
respect to the linear evolution, is shown in Fig.~\ref{fig:newexp}b
for an experimental dataset at
$\mu=\SI{0.12}{\milli\electronvolt}$. From Fig.~\ref{fig:newexp}a, it
can be seen that a faster expansion implies a smaller rotation angle
at long times, as shown in Fig.~\ref{fig:newexp}c by comparing the
linear to the non linear case. Although the deviation of the rotation
angle from the linear evolution becomes significantly appreciable only
at longer times, the global fit of expansion and rotation allows to
extract a reliable value for the prefactor $b$ in
Eq.~(\ref{eq:angle}). In Fig.~\ref{fig:newexp}d, the results for the
parameters $(a,b)$ obtained from the experiments shown in
Fig.~\ref{fig:dist} and Fig.~\ref{fig:newexp}, and from the numerical
simulations of GP dynamics are summarised for different chemical
potentials. The blue line corresponds to the curve $b = a_{0}/a$,
expected without the Magnus contribution. Both experiments and
simulations show a small, but measurable deviation from the blue
line. If the cores are in the parabolic region of the Gaussian
envelope the added velocity scales up linearly with $r$, without
disturbing the regular lattice shape. However the effect of additional
local density gradients, due to the presence of neighbouring vortices,
increase the distortion of the lattice from the regular shape, adding
noise to the measurements. Furthermore, the strength of the
interactions is responsible for sustaining on a longer time the
rigid-like behaviour of the lattice, dominated by the kinetic
energy. In the opposite limit of very small interactions, or waiting
enough time in the polariton evolution, this condition may cease to be
valid since the intervortex separation becomes comparable to the
healing length, and a new regime may arise, as reported in
\cite{Schweikhard}, and confirmed by our simulations \cite{suppl}.

We have shown that lattices of quantised vortices in 
out-of-equilibrium, untrapped quantum fluids exhibit a conformal
stretching and rotation, compensating the ballistic radial expansion
by a decreasing angular velocity. Interactions modify this picture: in
our repulsive case, a radial acceleration outwards increases the
inter-vortex separation and limits the rotation angle at long
times. The vortex lattice behaviour is compared to the quantized
circulation of the whole fluid, showing a Magnus-like effect as an
additional rotation of the vortex lattice with respect to the
fluid. These results show the crucial importance of having
experimental access to a well resolved space/time tracking of the
vortices in an expanding fluid, where nonlinear effect rapidly
weaken. Such high degree of control over the non-equilibrium dynamics
of interacting quantum fluids opens up the possibility to achieve
configurations with a larger vortex density and ad hoc, all-optical,
confining potentials. It is still an open question whether
turbulent-like regimes akin to what realised in other systems
\cite{Salort2010, Navon2016} will be within reach in exciton-polariton
fluids.

\begin{acknowledgements}
  We thank Marc E. Brachet, Natalia Berloff and Ricardo Carretero for
  useful discussions. The authors acknowledge the project PRIN
  ``Interacting Photons in Polariton Circuits INPhoPOL'' (MUR
  2017P9FJBS\_001), the project ``TECNOMED - Tecnopolo di
  Nanotecnologia e Fotonica per la Medicina di Precisione'', (MUR
  Decreto Direttoriale n. 3449 del 4/12/2017, CUP B83B17000010001) and
  the ``Accordo bilaterale CNR/RFBR (Russia) - triennio 2021-2023''.
\end{acknowledgements}

\cleardoublepage
\pagebreak
\clearpage 
\onecolumngrid

\setcounter{equation}{0}
\setcounter{figure}{0}
\setcounter{table}{0}
\setcounter{page}{1}
\makeatletter

\renewcommand{\bibnumfmt}[1]{[S#1]}
\renewcommand{\citenumfont}[1]{S#1}

\renewcommand{\theequation}{S\arabic{equation}}
\renewcommand{\thefigure}{S\arabic{figure}}
\renewcommand{\thetable}{S\arabic{table}}

\renewcommand\thepage{S\arabic{page}}

\phantomsection{}
\label{sec:Sinfo}
\noindent
\textbf{\large Dynamics of a vortex lattice in a expanding polariton quantum fluid \\Supplementary Material\\}

\normalsize



\vspace{1cm}

\twocolumngrid

\section{Methods}
\label{append:methods}
The semiconductor microcavity used in these experiments is a planar
one with 12 GaAs quantum wells embedded in two distributed Bragg
reflectors consisting of Al$_x$Ga$_{1-x}$As layers with aluminium
fractions of 0.2 and 0.95. As in most experiments of this kind, the
microcavity is kept at a cryogenic temperature of $\sim 5$ K. The high
Q-factor of this structure ($> 10^5$) results in a polariton
lifetime ($> 50$ ps), which allows the study of fairly long dynamics.

To shape the phase profile of the incoming laser beam, we used a
spatial light modulator (SLM), \textcolor{black}{a 1920x1080 pixels
  liquid crystals display with a surface area of approximately 2
  cm$^2$.  The effective refractive index seen by the incident wave is
  controlled by applying a voltage to the cells, which changes the
  orientation of the liquid crystals. The control of the birefringence
  of each pixel allows to spatially design the phase retardation of
  the reflected wave.  On top of the phase pattern of the vortex
  lattice, we used a blazed grating pattern, allowing us to block the
  zeroth-order reflection from the SLM that brings the non-modulated
  part of the laser.}
\begin{figure}[htbp]
    \centering
    \includegraphics[width=\columnwidth]{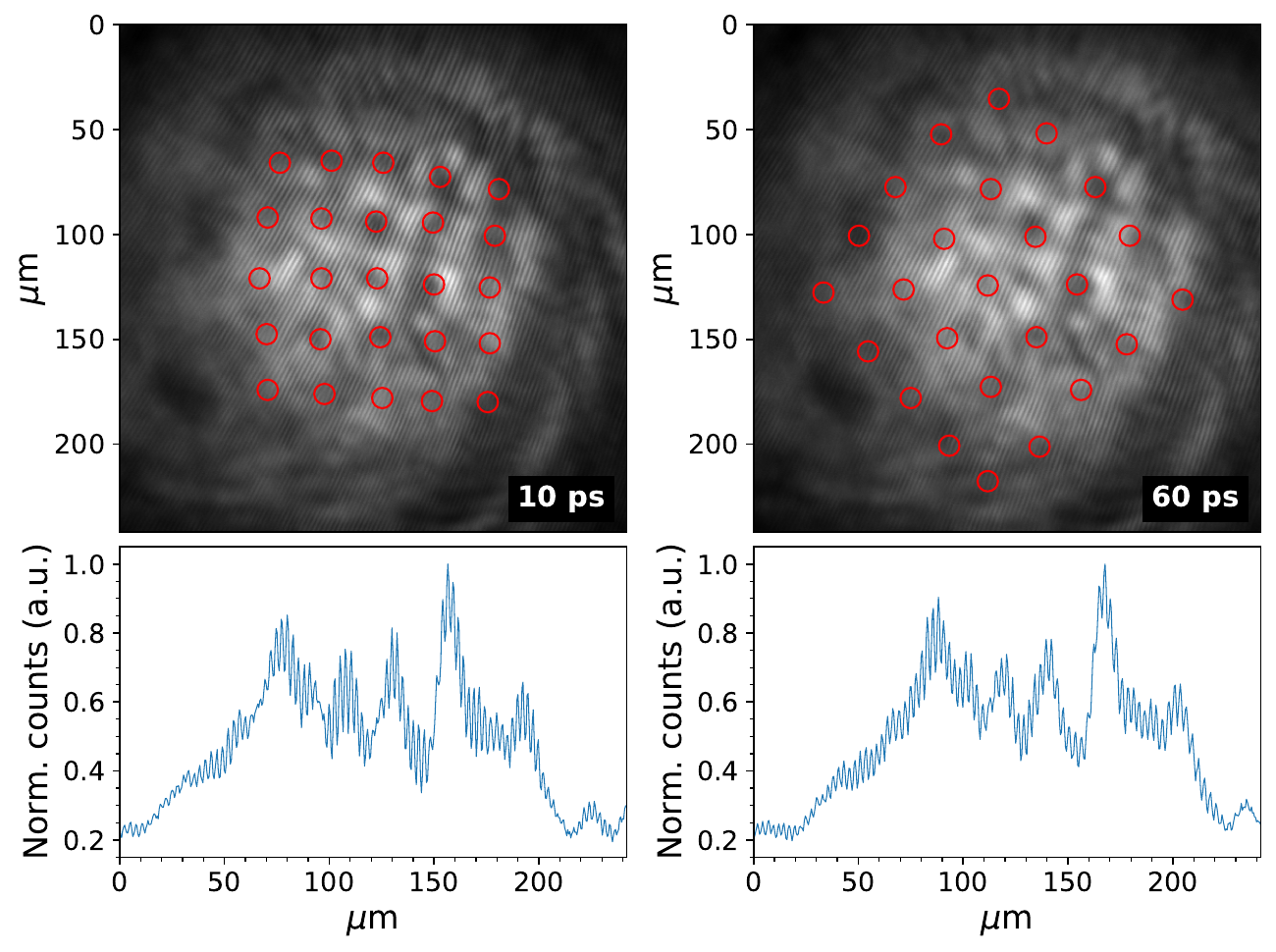}
    \caption{\textcolor{black}{Real space interferograms taken at 10 ps
        (left) and 60 ps (right), in which the vortices positions are
        recognisable by the characteristic ``forks'' highlighted by
        the red circles. On the bottom the intensity profiles for both
        cases; the signal to noise ratio decreases in time but still
        remains sufficiently high to accurately reconstruct the module
        and phase of the signal. Note that the background density in
        the interferograms is time-integrated: the temporal resolution
        results only from the interference with the reference.}}
    \label{fig:interferograms}
\end{figure}
A 80 MHz pulsed laser with a pulse duration of 2 ps is used to
resonantly excite the microcavity sample and directly create a
polariton fluid carrying the lattice of vortices. \textcolor{black}{The
  time dynamics are obtained by \textit{off-axis digital holography},
  a technique in which the emission from the sample is made interfere
  with a homogeneous homodyne reference beam. The resulting
  interferograms (Fig.~\ref{fig:interferograms}) are collected by a
  CCD and, by means of a digital Fast Fourier transform (FFT), we are
  able to extract information on both the amplitude and phase of the
  signal; by changing the reference delay, we are able to reconstruct
  the dynamics of the system with a time resolution of 1
  ps. Furthermore with the integration time of our CCD ($\sim$~1 ms)
  each interferogram is integrated over thousands of pulses, ruling
  out stochastic effects. The errors in the vortex separation and
  rotation angle are estimated at each time frame by the variance over
  the whole lattice. The fitting procedure includes the 95\%
  confidence levels obtained from these uncertainties in the
  estimation of the parameters in Eq.2 and 3, providing the standard
  errors for ``a'' and ``b''.}

The experimental setup is shown schematically in Fig.~1 in the main
text. The access to the phase of the macroscopic wavefunction allows
us to track the positions of the vortex cores by measuring phase jumps
of $2\pi$ in local circuitations, which represents a great advantage
with respect to atomic BECs or superfluid helium experiments.

\vspace{0.7cm}

\textcolor{black}{\section{Working conditions}}

\begin{figure}
    \centering
    \includegraphics[width=0.9\columnwidth]{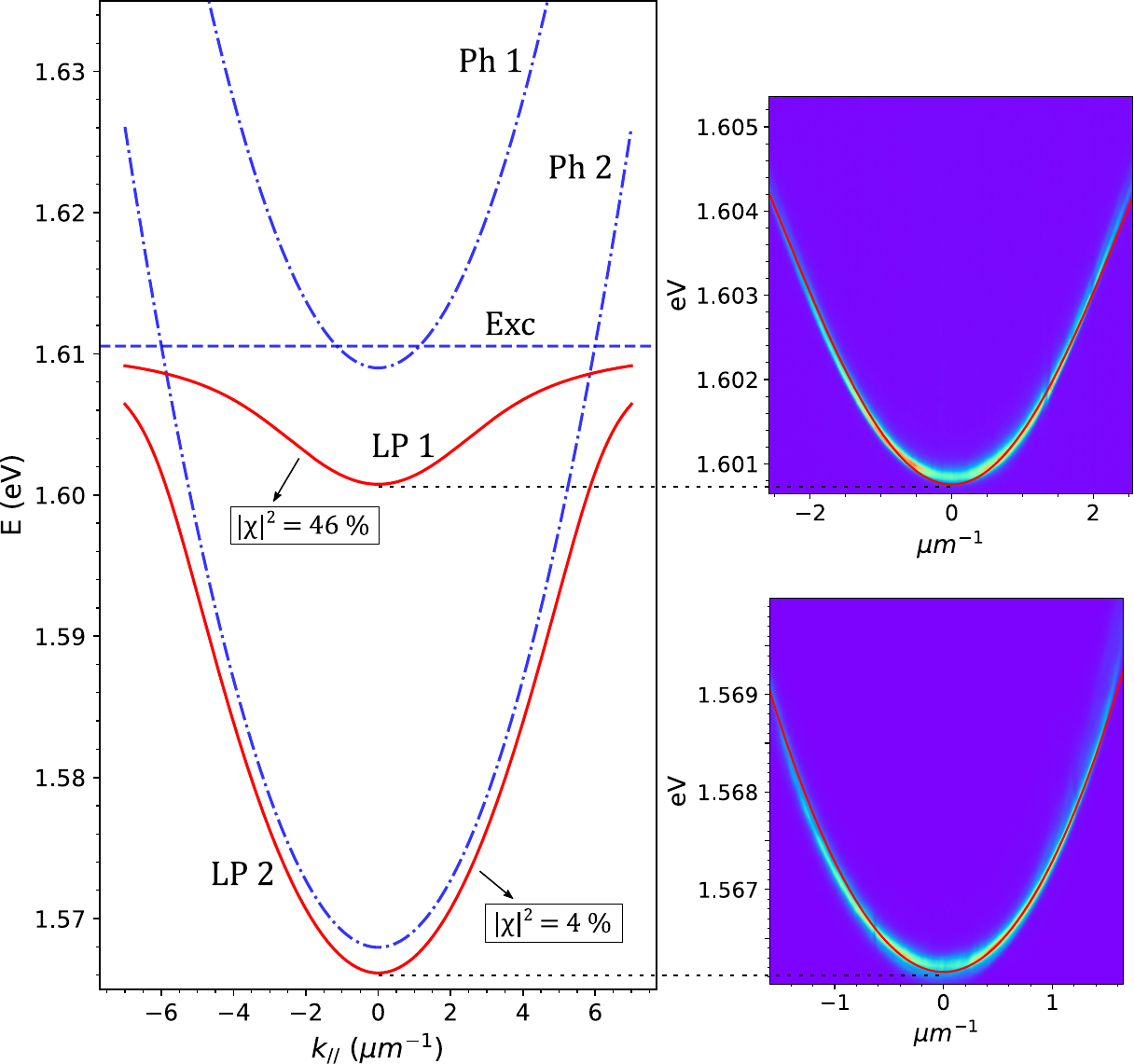}
    \caption{\textcolor{black}{Lower polariton branches (red lines)
        obtained from the best fit of the experimental dispersion on
        the right. The dashed line represents the energy of the
        exciton, while the dashed-dotted lines corresponds to the
        energy dispersions of the cavity photons for two different
        detunings.}}
    \label{fig:LP_disp}
\end{figure}

\textcolor{black}{When a microcavity mode has a sufficiently narrow
  linewidth, photons and excitons strongly couple, forming two new
  eigenstates
  $$\psi_{L}^U=\mathcal{X}\psi_{ex}\pm\mathcal{C}\psi_{ph}\,,$$ called
  exciton-polaritons, where $\mathcal{X}$ and $\mathcal{C}$ are the
  Hopfield coefficient and $|\mathcal{X}|^2 + |\mathcal{C}|^2 =
  1$. The square of this coefficients represents the
  photonic/excitonic fraction of polariton states. The thickness of
  the microcavity sample decreases moving from its center towards the
  edges due to the deposition technique. This thickness gradient
  causes the energy of the cavity photon to change throughout the
  sample, while the energy of the exciton remains the same. It is
  therefore possible to tune the detuning $\Delta E = E_{ph} -
  E_{ex}$, where $E_{ph}$ is the energy of the cavity photon at zero
  incident angle with respect to the sample surface ($k=0$), and
  $E_{ex}$ is the exciton energy. In the main text, we show results on
  two different sample positions corresponding to different detunings,
  as shown in Fig.~\ref{fig:LP_disp}. Changing the detuning
  corresponds to change the excitonic (and photonic) fraction of
  polaritons, effectively modifying their mass and the interaction
  strength. Indeed,
$$1/{m_{pol}} =|\mathcal{X}|^2/m_{x}+|\mathcal{C}|^2/m_{c}\,,$$ while
  $g_{pol}=|\mathcal{X}|^2g_{x}$, with $m_{pol}$, $m_x$ and $m_c$ the
  lower polariton mass, the exciton mass and the photon mass,
  respectively, and $g_{x}$ is the exciton-exciton interaction
  strength \cite{CiutiPRB}.}
\textcolor{black}{The excitonic fractions that we used are
  $|\mathcal{X}|^2 = 0.46$ (LP1) and $|\mathcal{X}|^2 = 0.04$ (LP2),
  corresponding to the data in Fig.3 and Fig.4 in the main text,
  respectively. The polariton effective mass close to the excitation
  frequency (i.e., close to $k=0$) can be estimated using the
  parabolic approximation
  $E(\mathbf{k})=E_0+(\hbar^2\mathbf{k}^2)/(2m)$. From the best fit of
  this equation, we obtain $m_{pol}= 3.69 \,10^{-5} \,m_e$ for LP1 and
  $m_{pol}=7.21\, 10^{-5} m_e$ for LP2, where $m_e$ is the free
  electron mass (or $m_{pol}=$ 0.22 ps$^2$meV/$\si{\micro}$m$^2$ for
  LP1 and $m_{pol}=$ 0.41 ps$^2$meV/$\si{\micro}$m$^2$ for LP2).}\\

\section{Configurations}
\label{append:configurations}
\begin{table}
\centering
\begin{tabular}{ C{0.7cm}||C{1.cm}|C{1.cm}|C{1.cm}|C{1.cm}|} 

 EXP & $d (\mu m)$ & $N$ & $Q$ & $m_{pol}$\\
\hline
A & 21 & 49 & -49 & 0.22 \\ 
\hline
B & 26 & 49 & -49 & 0.22 \\ 
\hline
C & 26 & 49 & 49 & 0.22 \\
\hline
D & 30 & 49 & -49 & 0.22 \\ 
\hline
E & 26 & 49 & $\sim$0 & 0.22 \\ 
\hline
F & 26 & 25 & -25 & 0.41 \\
\hline
\end{tabular}
\caption{Key parameters for each different configuration of the
  experiment. Experiments are labeled with letters in col.1; in col.2
  the initial inter-vortex separations; in col.3 the number of
  imprinted vortices; in col.4 the total topological charge: positive
  $Q$ cases have left circulation, negative $Q$ cases have right
  circulation; \textcolor{black}{in col. 5 the estimated polariton
    mass. Experiments from $A$ to $E$ are discussed in Fig.3 of the
    main body, while experiment $F$ is discussed in Fig. 4 of the main
    body.}}
\label{tab:conditions}
\end{table}

\textcolor{black}{We summarise some details of the different
  realizations of the experiment. Each case is labelled with a letter
  from A to F and corresponds to a change in either the inter-vortex
  initial separation ($d_0$), the number of vortices ($N$), the total
  topological charge ($Q$), or the detuning as summarized in
  Tab. \ref{tab:conditions}}.\\
\begin{figure}
   \centering
    \includegraphics[width=0.90\columnwidth]{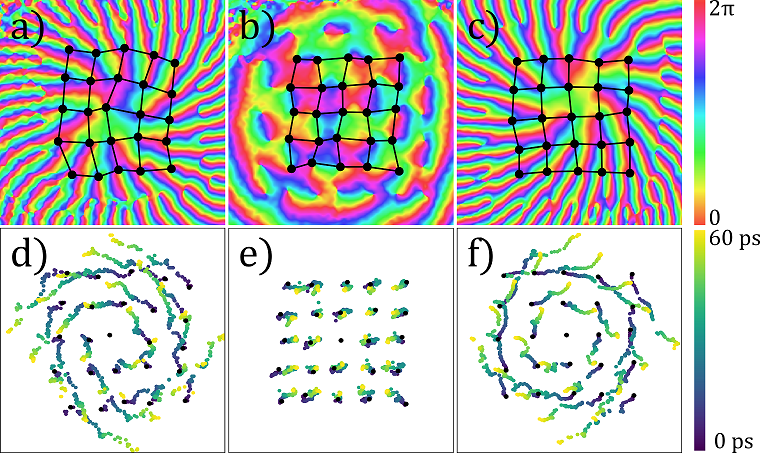}
    \caption{(\textbf{a}-\textbf{c}) Initial phase patterns for three
      different configurations, with same total number of vortices
      $N=49$ but different total charge, respectively $Q=49$,
      \textcolor{black}{$Q\simeq 0$} and $Q=-49$. Only the inner square
      sublattice of 25 vortex is highlighted. (\textbf{d}-\textbf{f})
      Trajectories of the inner 25 vortices over time for the
      corresponding configurations.}
    \label{fig:conditions}
\end{figure}
\begin{figure}
    \centering
    \includegraphics[width=0.7\columnwidth]{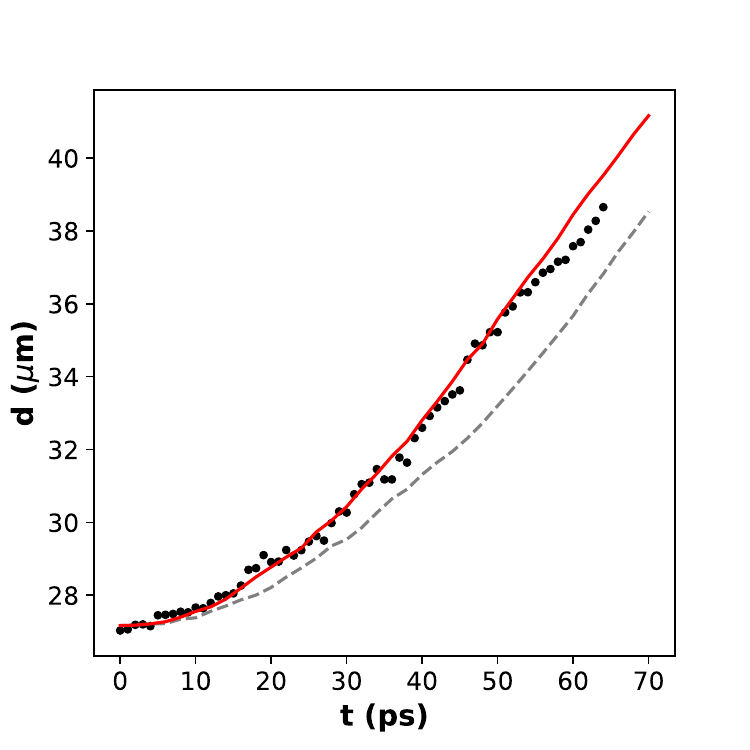}
    \caption{Evolution of the lattice radius. The black dots are the
      experimental data while the red line and the grey dashed line
      come from two simulations with and without a nonlinear term,
      respectively.}
    \label{fig:SI_expansion}
\end{figure}
The phase pattern of three lattices of $7\times7$ vortices, with total
charge equal to $Q=49$, \textcolor{black}{$Q\simeq 0$} and $Q=-49$ is
reported in Fig.~\ref{fig:conditions}(a-c), respectively. The inner
$5\times5$ sublattice is highlighted in the first row panels, with a
black dot on top of each phase singularity and a thin intervortex link
line in between them. The rotation of the fluid is measured from the
rotation of the lattice shown in Fig.~\ref{fig:conditions}(a-c) by
tracking the vortex positions in time. In
Fig.~\ref{fig:conditions}(d-e), the trajectories of the vortices, in a
time window of 60 ps, are shown for the three cases in
Fig.~\ref{fig:conditions}(a-c).\\ When the total topological charge is
almost null, as in Fig.~\ref{fig:conditions}(b,e), the quasi-neutral
lattice neither rotates nor expands; on the contrary, it slightly
shrinks because of vortex-antivortex attraction. It is possible to
spot both in the time integrated 2D trajectories on the map, and also
in the corresponding intervortex distance, or lattice cell parameter,
which is derived as a mean among all the first neighbors
pairs. Indeed, for the $Q\approx 0$ lattice (green dots), there is a
continuously decreasing intervortex distance in the observed
dynamics. The polariton lifetime is not long enough to observe
vortex-antivortex annihilation in the experiments, while this effects
can be measured in numerical simulation at longer time (not
shown). For both cases $Q=\pm49$, whose initial phase maps are shown
in Fig.~\ref{fig:conditions}(a,c) and $xy-$trajectories are shown in
Fig.~\ref{fig:conditions}(d,f), the lattice does evolve under the
combined effect of a rotation and expansion movement.\\

\section{Feynman-Onsager relation}

The velocity circulation around a multiply connected region $A$
enclosing $N$ unitary vortices of total charge $Q$ is quantised and
can be written as
\begin{equation}
\oint_{\partial{A}} {\bf v} \cdot \mathrm{d}{\bf l} = Q\, \frac{\it h}{m}\,,
  \label{eq:circ}
 \end{equation} 
where the superfluid velocity is defined as 
\begin{equation}
{\bf v}({\bf r},t)=\frac{\hbar}{m}{\bm\nabla}\varphi({\bf r},t),
\end{equation}
with $m$ the polariton mass and $\varphi({\bf r},t)$ the phase of the
macroscopic condensate wavefunction. When the area $A$ is uniformly
covered by $Q$ vortices with spacing $d$, we have that
\begin{equation}
\int_{A} ({\bm \nabla} \times {\bf v})\cdot \mathrm{d}{\bf A} = 2 Q \Omega d^2,
  \label{eq:rot}
\end{equation}
which defines an overall vorticity associated to the angular rotation
velocity $\Omega$ of the superfluid. By the Stokes theorem, it follows
the Feynman-Onsager relation,
\begin{equation}
\frac{\it h}{2m} = \Omega d^2 \,.
\label{eq:feynman}
\end{equation}

\section{Direct Numerical Simulations and Fitting}
\label{append:fitting}
Polariton-polariton interactions are the key element that makes the
difference between a pure optical beam and a polariton beam. This is
clear in Fig.~\ref{fig:SI_expansion}, where we show that for a
simulation (red solid line) to match the experimental evolution of the
lattice radius (black dots) it is necessary to consider a nonlinear
term \textcolor{black}{($\mu = ng \simeq 0.05$ meV)}. In constrast, a linear
expansion (grey dashed line) is much slower.
\begin{figure}
    \centering
    \includegraphics[width=0.9\columnwidth]{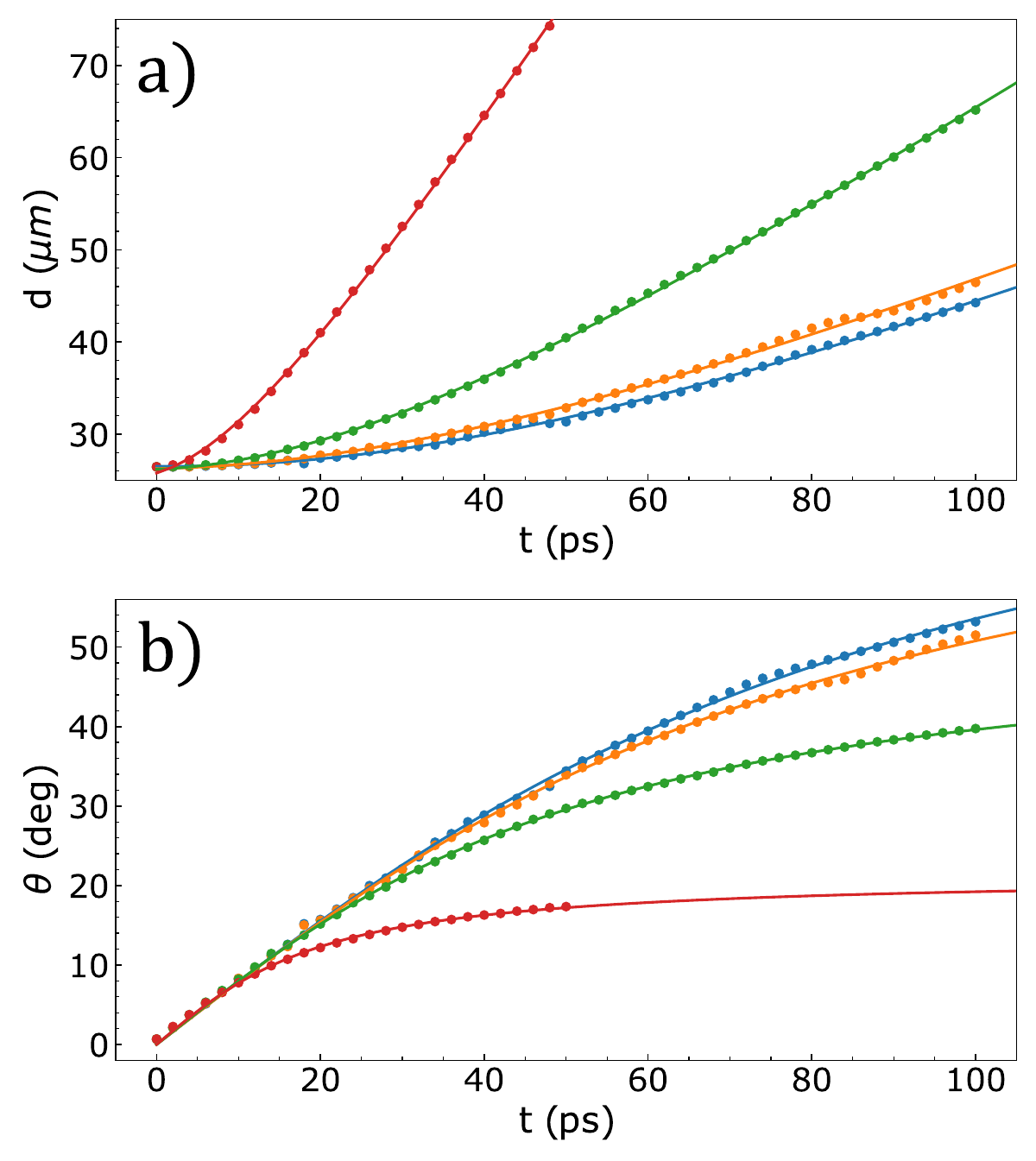}
    \caption{\textcolor{black}{Geometrical parameters of the vortex
        lattice obtained from the numerical simulation of the GP
        equation, at varying the nonlinear coupling strength $g$
        (initial density $n_0$ is the same in all simulations),
        expressed in meV$\si{\micro}$m$^2$: blue $g \simeq 0$
        (i.e. the linear case), orange $g=5.\,\mathrm{10}^{-4}$, green
        $g=5. \,\mathrm{10}^{-3}$, red
        $g=5.\,\mathrm{10}^{-2}$. (\textbf{a}) The mean inter-vortex
        separation $ d(t)$, computed as the average over the first
        neighbors distances in the $5 \times 5$ sublattice
        (dots). Solid lines are the fit of Eq.~2 in the main text,
        with different value of the parameter $a$. (\textbf{b}) The
        lattice angles of orientation $\theta$ (dots) and their fits
        (solid lines) with Eq.~3 in the main text. The parameter $b$,
        representing the limit angle of rotation, decreases as the
        nonlinearity is increased.}}
    \label{fig:expansion}
\end{figure}

\textcolor{black}{Compared to the pure optical case, if we add
  interactions to the polariton condensate, the expansion is faster,
  given that the interaction energy at $t=0$ is released into kinetic
  energy during the initial evolution of the superfluid. Moreover, the
  nonlinear interaction plays the role of a radial force, bending
  vortex trajectories at short times. \\In Fig.~\ref{fig:expansion},
  we compare the time evolution of the system, numerically integrating
  the GP equation, with four different values of the nonlinear
  interaction strength $g$, but the same initial state with a regular
  lattice of $Q=49$, $d_0=$ 26 $\si{\micro}$m and density $n_0\simeq
  100/\si{\micro}$m$^2$. In the linear case, each vortex would move
  along a straight line, while the nonlinear repulsion (taken as a
  purely radial force) is responsible for a deviation in the
  trajectory due to the earlier onset of a radial component of the
  velocity proportional to the interaction strength. The deflection
  asymptotically vanishes as it does the polariton density, due to
  both losses and expansion.}\\
\section{Numerical methods}
\label{append:DNS}

The spatio-temporal evolution of the vortex lattice within the
polariton fluid is described in terms of the two-dimensional
Gross-Pitaevskii model equation for the condensate macroscopic
wavefunction $\Psi({\bf r},t)$,
\begin{equation}
\label{eq:GP}
    \it{i}\hbar \frac{\partial \Psi({\bf r},t)}{\partial t} = \left[- \frac{\hbar^2 }{2m_{pol}}\,\nabla^2 + g |\Psi({\bf r},t)|^2 - \frac{i \hbar \Gamma}{2}\right] \Psi({\bf r},t)\,,
\end{equation}
where $m_{pol}$ is the effective mass of the lower polariton branch, $g$ is
the polariton-polariton interaction strength, and $\Gamma$ is the
lower polariton loss rate \cite{Carusotto2013}. Numerical simulations
of the equation above are performed using a pseudo-spectral code, with
a standard split-step Fourier method for the kinetic term
\cite{MSTT04}, and a Runge-Kutta method for the non-linear term solved
in real space. The physical domain is $[-L:L]$, and is
discretised on an regular square $N_X \times N_Y$ grid. De-aliasing
with the $2/3$ rule (i.e., at $k_{max}=\frac{2}{3}\frac{N_X}{2}$) is
applied \cite{GO77}.\\ All runs are initialised with the condensate
wavefunction $\Psi({\bf r})$, such that \cite{FS01}
\begin{eqnarray}
\label{eq:IC}
&&\Psi({\bf r}) = \sqrt{n_0}\displaystyle\prod_{j=1}^{N} \zeta({\bf r}_j/\xi)\,e^{\pm i \phi},\\
&&\zeta ({\bf r_0}/\xi) = \frac{\sqrt{(x-x_0)^2 + (y-y_0)^2}}{\sqrt{\xi^2/\Lambda^2 + (x-x_0)^2 + (y-y_0)^2}}\,.
\end{eqnarray}
If equispaced, this corresponds to a regular, square lattice with $N$
single-charge vortices, each vortex being centered at ${\bf
  r}_j=(x_j,y_j)$; $n_0$ is the uniform polariton density amplitude;
the parameter $\Lambda=0.8249$ controls the slope of vortex solution
in the deep core of the vortex, and $\xi=\hbar/\sqrt{g\,m_{pol}\,n_0}$
is the healing length estimated on the initial uniform condensate
density. For the scaling ansatz (\ref{eq:IC}) to be valid, the lattice
spacing ($d=|{\bf r}_i - {\bf r}_j|$) has to be much larger than the
healing length $\xi$. \\ We performed different series of runs by
varying the number of vortices $N$, the total charge of the lattice
$Q$, the initial value of the density $n_0$, the lattice spacing
$d(t=0)$, the grid resolution $N_X$, and the non-linear coupling
strength $g$. In the paper, we limit our discussion to two series of
runs. \\Series I related to experimental data of Fig. 3 of the main
text: the total number of vortices $N=49$, and the topological charge
$Q= (+49, -49, 0)$, the number of grid point $N_X \times N_Y=2048^2$;
the size of the domain $L=350 \si{\micro}$m; the initial lattice
spacing $d(0)= 26 \si{\micro}$m. The lower polariton mass, estimated
from the sample properties, is $m_{po} =0.22$
ps$^2$meV/$\si{\micro}$m$^2$; the polariton-polariton interaction
strength is spanned over the values $g = 5\cdot[10^{-7}, 10^{-4},
  10^{-3}, 10^{-2}]$ $ \SI{}{\meV\micro\meter^2}$; the time step
$\delta t=0.025$ ps and the run duration $T$ between $T=(100,200)$
ps. Moreover for the runs here discussed we fixed the loss rate
$\Gamma=0$. Finally, we have one free parameter to match experimental
conditions, that is the initial uniform density $n_0$: for the runs
here discussed it is $n_0=100/\si{\micro}$m$^2$, so that the chemical
potential is in the range $\mu=gn=[5.10^{-5};5.0]$ meV. With this set
of parameters, the grid resolution $dx = 2 L/N_X =0.3 \si{\micro}$m is
always comparable or smaller than the healing length, i.e. $\xi_0 \in
[0.1,14] \si{\micro}$m for $g$ in the explored
range.\\ \textcolor{black}{Series II related to experimental data of
  Fig. 4 of the main text: $N = |Q| = 25$; the number of grid points
  $N_X \times N_Y=1024^2$; the size of the domain $L=315
  \si{\micro}$m; the initial lattice spacing $d(0)= 26.5
  \si{\micro}$m. The lower polariton mass, estimated from the sample
  properties, is $m_{pol} =0.41$ ps$^2\mathrm{meV}/\si{\micro}$m$^2$;
  the chemical potential, based on the sample properties, is in the
  range $\mu = g n \simeq [0.01;0.6] \mathrm{meV}$ (see Fig.4 in the main
  text); the time step $\delta t=0.02$ ps and the run duration is
  $T=100$ ps.} \\Following the experiments, numerical simulations are
performed in the absence of a confining potential: this implies that
the condensate expands, and the kinetic energy ($E_{kin} =
(\hbar^2/2m) \int |\nabla \Psi|^2\, d^2 {\bf r}$) grows in time at the
expenses of the interaction energy ($E_{int} = g/2 \int |\Psi|^4\, d^2
{\bf r}$).

\end{document}